\documentclass[8pt]{article}

\usepackage{amsmath,amssymb,amsthm,graphicx,latexsym}

\newcommand{\ed}{\end{document}}
\newcommand{\beq}{\begin{equation}}
\newcommand{\eeq}{\end{equation}}

\begin{document}
\begin{center}
\large{\textbf{\textbf{GUP-based and Snyder Non-Commutative Algebras,
Relativistic Particle models, Deformed Symmetries and Interaction: A Unified 
Approach}}}\\
\end{center}
\begin{center}
Souvik Pramanik\footnote{E-mail: souvick.in@gmail.com} and
Subir Ghosh\footnote{E-mail: sghosh@isical.ac.in} \\
Physics and Applied Mathematics Unit, Indian Statistical
Institute\\
203 B. T. Road, Kolkata 700108, India \\
\end{center}
\vspace{0.5cm}

\textbf{Abstract:} We have developed a unified scheme for studying
Non-Commutative algebras
based on Generalized Uncertainty Principle (GUP) and  Snyder form in a
relativistically
covariant point particle Lagrangian (or symplectic) framework. Even though the
GUP
based algebra and   Snyder algebra are very distinct, the more involved latter
algebra
emerges from an approximation of the Lagrangian model of the former algebra.
Deformed
Poincare generators for the systems that keep space-time symmetries of the
relativistic
particle models have been studied thoroughly.

From a purely constrained dynamical analysis perspective the models studied here
are very
rich and provide insights on how to consistently construct approximate models
from the
exact ones when non-linear constraints are present in the system.

We also study dynamics of the GUP particle in presence of external
electromagnetic field.

\section{Introduction:}
Operatorial forms of Non-Commutative (NC) phase space structures, of the generic
form,
\beq \{x_i,p_j\}=\delta_{ij}(1 + f_1(\mathbf p^2)) +f_2(\mathbf p^2) p_i
p_j,$$$$
\{x_i,x_j\}=f_{ij}( \mathbf p),~\{x_i,x_j\}=g_{ij}(\mathbf p);~~i=1,2,3,
\label{xp}\eeq
have created a lot of interest in recent years due to their potential
application
in generating Generalized Uncertainty Principle (GUP), the latter being
compatible
with String Theory or Quantum Gravity expectations of the presence of a minimum
length scale or a maximum momentum scale or both. $\beta $ is treated as a small
parameter with ${\sqrt{\beta }}$ being the measure of a minimum length scale.
The
argument goes as follows: resolution of position coordinate to an arbitrary
precision
can lead, via (Heisenberg) canonical uncertainty principle, to such a large
accumulation
of momentum or energy density that the latter can appreciably alter the
space-time
metric. Calculations from String Theory perspective \cite{str} also suggest a
minimum
length scale in the form of minimum position uncertainty. This is the possible
origin
of GUP. The pioneering works proposing consistent NC algebras  with a similar
motivation,
in a non-covariant framework, were by Kempf \cite{kem},
\beq \{x_i,p_j\}=\delta_{ij}(1 +\beta {\mathbf p}^2)  +\beta ' p_i p_j ,~
\{x_i,x_j\}=(\beta'-2\beta )(x_ip_j-x_jp_i) ,~ \{p_i,p_j\}=0
.\label{ppcommutator}\eeq
by Kempf, Mangano and Mann \cite{kem1}
\beq \{x_i,p_j\}=\delta_{ij}(1 +\beta {\mathbf p}^2)  ,~\{x_i,x_j\}=-2\beta
(x_ip_j-x_jp_i) ,~ \{p_i,p_j\}=0, \label{ppcommutator1} \eeq
and by Kempf and Mangano \cite{kem2},
\beq \{x_i,p_j\}=\frac{\beta {\mathbf p}^2\delta_{ij}}{\sqrt{(1+2\beta {\mathbf
p}^2)}-1}+\beta p_i p_j ,~ \{x_i,x_j\}=0 ,~ \{p_i,p_j\}=0
.\label{ppcommutator2}\eeq
We are restricting ourselves to the classical counterpart of the commutators but
the
results derived here are applied to quantum commutators as well. The first work
on
operatorial NC algebra was by Snyder \cite{sn} with the structure same as that
of
(\ref{ppcommutator1}). In fact (\ref{ppcommutator}) \cite{kem} and
(\ref{ppcommutator1})
\cite{kem1} are a generalized form of \cite{sn} and can be reduced to the Snyder
form
of NC \cite{sn} by suitable choice of parameters, as discussed by \cite{q}.
However,
we will focus principally on the  algebra (\ref{ppcommutator2}) \cite{kem2},
(subsequently
referred as KM) since it is structurally the simplest as the coordinates and
momenta
commute among themselves respectively.

The work reported in the present paper can be divided in to three parts which
are
inter connected. We work with a relativistically covariant generalization of the
algebra (\ref{ppcommutator2}). In the first part we derive a generalized point
particle Lagrangian with a non-canonical symplectic structure that is equivalent
to (\ref{ppcommutator2}). Clearly from a physics point of view this type of an
intuitive  particle picture is very useful and appealing since we can see how it
differs from the conventional relativistic point particle. Primarily it is
essential
in studying the dynamics of such particles (in Hamiltonian framework) that can
reveal unique features of such particles. This can act as a precursor to field
theories in such non-canonical space. It is not possible to obtain such
information
only from the phase space algebra which is essentially kinematical in nature.
 Similar point particle symplectic
formalisms have been adopted for other forms of operatorial NC algebras such as
$\kappa$-Minkowski algebra \cite{sg1,others}, relevant in Doubly Special
Relativity
framework \cite{am} or Very Special Relativity algebra \cite{sg2}, proposed in
\cite{gl}.

The point particle scheme is crucial for the second part of our work. This deals
with the richness and intricacies of Dirac formalism \cite{dir} when applied to
non-linear constraints, (i.e. constraints consisting of non-linear terms in
phase
space variables), which are necessary to induce operatorial phase space algebras
as Dirac Brackets. In an explicit way we will show that the simpler algebra
(\ref{ppcommutator2}) can  be ``reduced'' to the more complicated Snyder form
Snyder \cite{sn, kem1}. ``Simple'' and ``complicated''  refer to $\{x_\mu,x_\nu
\}$
being zero or non-zero respectively. It is clearly revealed how approximate
forms
of the Lagrangian, related to (\ref{ppcommutator2}) \cite{kem2}, with terms up
to
at least $O(\beta ^2)$  can reproduce the algebra (\ref{ppcommutator1})
\cite{kem,kem1}.
Furthermore, a simpler Lagrangian with two parameters $\beta $ and $\beta '$ is
also
provided that gives rise to the Snyder algebra. This strongly brings in to fore 
a
point that it is always advisable to impose approximations in a theory at the
Lagrangian level and then go on to compute the symplectic structure rather than
truncate the exact symplectic structure directly. In the latter procedure it is
natural to encounter consistency problems in the algebra,(such as violation of
Jacobi identity), leading to incorrect conclusions. Furthermore, the elegant
connection
between qualitatively different algebras, (such as \cite{kem2} and
\cite{kem,kem1,sn},
derived at different stages of approximation of the Lagrangian, will be lost.
The
comment below is relevant in this context.

Another very important aspect of point particle framework is that one can
introduce interactions in a consistent way. This is the topic of the third part.
It is
possible that  interactions can bring out certain interesting or even unphysical
features (if there are any) of the generalization which do not show up in the
free
particle context. (For a recent work in this regard see for example \cite{sg3}
where
$\kappa$-Minkowski particles are subjected to electrodynamic interactions.)

A covariantization of \cite{kem1,sn} was performed in \cite{q} for the
\cite{kem1} case:
\beq \{X^\mu,P^\nu\}=\delta^{\mu\nu}(1+\beta P^2)+\beta'P^\mu
P^\nu\label{XmuPnucom},
~ \{P^\mu,P^\nu\}=0,$$$$
\{X^\mu,X^\nu\}=\frac{(2\beta-\beta')+(2\beta+\beta')\beta P^2}{1+\beta
P^2}(P^\mu X^\nu-P^\nu X^\mu)\label{XmuXnucom}
\eeq
where $\mu,\nu=1,2,3$. However, the subsequent analysis to $O(\beta )$ is
doubtful to say the least since as the authors themselves admit in \cite{q}
the Jacobi identity is maintained by the linearized algebra,
\beq \{X_\mu,P_\nu\}=(\delta_{\mu\nu}(1+\beta P^2)+2\beta P_\mu P_\nu),$$$$
 \{P_\mu,P_\nu\}=\{X_\mu,X_\nu\}=0 \label{PPXXcom} \eeq
{\it{only}} to $O(\beta  )$. Since the violation of Jacobi is of the following
operatorial form,
$$J(X_\mu,X_\nu,P_\lambda)=\{X_\mu,\{X_\nu,P_\lambda\}\}+\{P_\lambda,\{X_\mu,
X_\nu\}\}+\{X_\nu,\{P_\lambda,X_\mu\}\},$$
which implies
\beq J(X_\mu,X_\nu,P_\lambda)=4\beta^2P^2(\delta_{\nu\lambda}P_\mu-\delta_{\mu
\lambda}P_\nu), \label{jacobi} \eeq
it is possible that the expectation value of the RHS of (\ref{jacobi}) becomes
large rendering the claim, that $O(\beta ^2)$ contribution is always small,
meaningless. Exact validity of Jacobi identity is imperative for the phase space
algebra. Furthermore, due to this violation of Jacobi, there can not be any
point
particle interpretation of this NC symplectic structure since, (indeed, from our
perspective), the NC structures appear as Dirac Brackets which always preserve
Jacobi identity \cite{dir}. We have also constructed deformed Poincare
generators
that generate proper translations and rotations of the variables.

The paper is organized as follows. In Section 2 we propose the covariant GUP
and develop the point particle Lagrangian corresponding to it and study the
space-time symmetry properties of this novel particle model. In Section 3 we
show how different forms of covariantized Snyder algebra can be generated from
approximations of GUP particle model. As a bonus we also obtain point particle
Lagrangians for these Snyder algebras. In Section 4 we discuss the GUP particle
interacting with $U(1)$ gauge fields.
The paper is  concluded in Section 5  with
a summary of our work and future directions. Some computational details are
provided in the appendices at the end.
\section{Covariantized GUP and the Point Particle}
We begin by positing  covariantized form of the NC algebra proposed in
\cite{kem2} in 3+1-dimensions, with a Minkowski metric $g_{00}=-g_{ii}=1$,\\
$$ \{x_\mu,p_\nu\}=-\frac{\beta p^2g_{\mu\nu}}{\sqrt{(1+2\beta p^2)}-1}-\beta
p_\mu p_\nu \equiv -\Lambda g_{\mu\nu}-\beta p_\mu p_\nu ,$$
\beq\{x_\mu,x_\nu\}=0 ,~  \{p_\mu,p_\nu\}=0, \label{ppc}\eeq
where $\Lambda =\frac{\beta p^2}{\sqrt{(1+2\beta p^2)}-1}$. For the spatial
sector this reduces to
$$\{x_i,p^j\}= \Lambda \delta ^{ij}+\beta p^i p^j, $$
similar as in \cite{kem2} with a mismatch in the value of $\Lambda $ (since
$\mathbf p^2$ has been replaced by $p^2$). We would like to interpret the above
relations (\ref{ppc}) as Dirac Brackets derived from a constrained symplectic
structure. In some sense we are actually moving in the opposite direction of the
conventional analysis where the computational steps are
$$Lagrangian~\rightarrow ~Constraints~\rightarrow ~ Dirac ~Brackets $$ or
equivalently
$$ Symplectic ~Structure~\rightarrow ~Symplectic~Matrix ~\rightarrow ~
Symplectic ~Brackets . $$ The Dirac brackets and symplectic brackets turn out to
be same. In our case the situation is reversed and our path of analysis will be
$$ Dirac ~Brackets~\rightarrow ~Constraints~\rightarrow ~Lagrangian $$ or
$$ Symplectic ~Brackets~\rightarrow ~Symplectic~ Matrix~\rightarrow ~Lagrangian
.$$
The procedure is the following. The generic form of a Symplectic Bracket (SB) is
of the form
\beq \{f,g\}_{SB}=\Gamma ^{\mu\nu}_{ab}\partial_{a,\mu}f\partial_{b,\nu}g,
\label{sb1} \eeq
where $\partial_{a,\mu}=\frac{\partial}{\partial\eta _a^\mu},~\eta_1^\mu
=x^\mu,~ \eta_2^\mu =p^\mu $. This Symplectic Matrix also appears in the Dirac
Brackets as
\beq \{f,g\}_{DB}=\{f,g\}-\{f,\Phi_a^\mu\}\Gamma ^{\mu\nu}_{ab}\{\Phi_b^\nu,g\},
\label{db1} \eeq
where $\Phi_a^\mu$ are a set of Second Class Constraints \cite{dir} (see
appendix
for a brief description of the Dirac procedure). Inverse of the $\Gamma $-matrix
provides the constraint algebra
\beq \Gamma _{\mu\nu}^{ab}=\{\Phi^a_\mu,\Phi_a^\nu\}. \label{cc} \eeq
Indeed there is no unique way but from the nature of the constraint matrix one
can make a judicious choice of the constraints and subsequently guess a form of
the Lagrangian. We do not claim the Lagrangian derived in this way is unique,
(in fact there might be more than one Lagrangians generating identical Dirac
Brackets), but at least one can easily check that the derived Lagrangian yields
the Dirac Brackets that one posited at the beginning.

Comparison with (\ref{ppc}) allows us to identify
\begin{equation}
 \{x^\mu,x^\nu\}\equiv \Gamma^{\mu\nu}_{11}=0,~\{p^\mu,p^\nu\}\equiv
\Gamma^{\mu\nu}_{22}=0,$$$$
\{x^\mu,p^\nu\}\equiv \Gamma^{\mu\nu}_{12}=-(\Lambda g^{\mu\nu}+\beta p^\mu
p^\nu )\equiv -\Gamma^{\mu\nu}_{12} .
\label{gam}
\end{equation}
The Symplectic Matrix is
\begin{equation} \Gamma_{ab}^{\mu\nu}=
\left[ {\begin{array}{cc}
0 & -(\Lambda g^{\mu\nu}+\beta p^\mu p^\nu )\\
(\Lambda g^{\mu\nu}+\beta p^\mu p^\nu ) & 0 \\
\end{array} }\right].
\end{equation}
 The inverse
matrix is computed and it turns out to be the commutator matrix,
\begin{equation} \Gamma^{ab}_{\nu\lambda}=
\left[ {\begin{array}{cc}
 0 & (\frac{g_{\nu\lambda}}{\Lambda}-\frac{\beta p_\nu
p_\lambda}{\Lambda^2{\sqrt{1+2\beta p^2}}}) \\
 -(\frac{g_{\nu\lambda}}{\Lambda}-\frac{\beta p_\nu
p_\lambda}{\Lambda^2{\sqrt{1+2\beta p^2}}}) & 0 \\
 \end{array} }\right].
\end{equation}
It is convenient to work in the first order formalism where both $x_\mu$ and
$p_\mu$ are treated as independent
variables with the conjugate momenta, $\pi_\mu^x=\frac{\partial
L}{\partial\dot{x^\mu}},~\pi_\mu^p=\frac{\partial
L}{\partial\dot{p^\mu}} $ with two decoupled canonical algebra
$$\{x_\mu,\pi_\mu^x \}=-g_{\mu\nu},~ \{p_\mu,\pi_\mu^p \}=-g_{\mu\nu}.$$
We propose the following set of constraints:
\beq \Phi_\mu^1=\pi_\mu^x \approx 0,\label{phi1}\eeq
\beq \Phi_\mu^2=\pi_\mu^p+\frac{x_\mu}{\Lambda}-\frac{\beta(x
p)p_\mu}{\Lambda^2\sqrt{1+2\beta p^2}} \approx 0.\label{phi2}\eeq
Thus we recover  explicit expressions for the momenta:
\beq \pi_\mu^x=\frac{\partial L}{\partial\dot{x_\mu}}=0,~
\pi_\mu^p=\frac{\partial
L}{\partial\dot{p_\mu}}=-\frac{x_\mu}{\Lambda}+\frac{\beta(x
p)p_\mu}{\Lambda^2\sqrt{1+2\beta p^2}}.\label{pimup}\eeq
Finally we can write down the cherished form of the point particle Lagrangian in
the first order form as,
\beq L=-\frac{(x\dot{p})}{\Lambda}+\frac{\beta(x
p)(p\dot{p})}{\Lambda^2\sqrt{1+2\beta p^2}}+\lambda (f(p^2)-m^2),\label{lag}\eeq
where $\lambda $ is a Lagrange multiplier. We have included a mass-shell
condition $f(p^2)-m^2=0$ where $f(p^2)$ denotes an arbitrary function that needs
to fixed. This is done from hindsight since we will show below that this
structure
will be invariant under modified Lorentz generators. This particle model is one
of
our major results.

First of all it is straightforward to check that from the Lagrangian, through
conventional Dirac Hamiltonian analysis of non-commuting constraints (or Second
Class Constraints, as they are termed in literature),  the algebra (\ref{ppc})
can be derived as Dirac Brackets.

Let us check how the new Lagrangian fares as regards the conventional space-time
symmetries, in particular translation and
generalized rotation (i.e. spatial rotation and boosts). From the NC algebra
(\ref{ppc}) it is clear that the momentum
$p^\mu$ can not play the role of Translation generator because of the anomalous
translation of $x^\mu$,
\beq \delta x_\mu=\{x^\mu ,(\sigma p)\}=-(\Lambda\sigma_\mu+\beta(\sigma p)p_\mu
),\label{tr}
\eeq
where $\sigma_\mu$ is the translation parameter. But a straightforward
generalization of a transformation of
variables proposed in \cite{kem2} shows that $\frac{p^\mu }{\Lambda }$ can act
as the
true Translation generator,
\beq \delta x_\mu=\{x^\mu ,\sigma_\mu (\frac{p^\mu }{\Lambda })\}=-\sigma_\mu.
\label{tr1}
\eeq
In a similar way  the Lorentz generators also get modified to
\beq j_{\mu\nu}=\frac{1}{\Lambda}(x_\mu p_\nu-x_\nu p_\mu),\label{jmunu}\eeq
such that correct transformation of the degrees of freedom are reproduced,
\beq [j_{\mu\nu},p_\lambda]=g_{\mu\lambda}p_\nu-g_{\nu\lambda}p_\mu ,~
 [j_{\mu\nu},x_\lambda]=g_{\mu\lambda}x_\nu-g_{\nu\lambda}x_\mu .
\label{jmunuxlamdacom}\eeq
Indeed $j_{\mu\nu}$ obeys the correct Lorentz algebra,
\beq
\{j_{\mu\nu},j_{\alpha\beta}\}=g_{\mu\alpha}j_{\nu\beta}-g_{\mu\beta}j_{
\nu\alpha}
-g_{\nu\beta}j_{\alpha\mu}+g_{\nu\alpha}j_{\beta\mu}.\label{lor}
\eeq
Let us fix the function $f(p^2)$ in the mass shell condition. Since
$\{j_{\mu\nu},p^2\}=0$ any funtion
of $p^2$ is Lorentz invariant but keeping Translation invariance in mind, a more
natural choice would be
$f(p^2)\rightarrow (\frac{p_\mu }{\Lambda})^2$ leading to a modified mass shell
condition
$ (\frac{p_\mu }{\Lambda})^2 -m^2=0$. However
this actually simplifies to $p^2=M^2,M=m/(1-\frac{\beta m^2}{2})$.

Quantization of this GUP particle can be carried through in the conventional way
by gauge fixing the
reparameterization invariance.

The above analysis demonstrates that we have developed a consistent and
relativistically covariant framework
to represent a generalized point particle living in an NC phase space compatible
with GUP.
\section{Approximations leading to other algebras}
 As we have explained at the beginning,
approximating the full NC
algebra (\ref{ppc}) directly is not the proper way to derive an effective
$O(\beta )$ corrected dynamical
system since, in particular with operatorial NC algebras, there is always a
drawback that Jacobi identities
night be violated. The correct way is to to approximate the system at the level
of the Lagrangian because then we
are assured that the $O(\beta )$ corrected NC brackets will also satisfy the
Jacobi identities. (Details of the Dirac Bracket calculations are provided in
the appendices.)
\vskip.4cm
{\bf{$O(\beta )$ {\it{results}}:}}  With $\Lambda =1+\frac{1}{2}\beta p^2 +
O(\beta
^2)$ the Lagrangian (\ref{lag})
yields $L_{(1)}  $ (without the mass-shell condition) to $O(\beta )$:
\beq L_{(1)}= -(x\dot{p})(1-\frac{1}{2}\beta p^2)+\beta(x p)(p\dot{p}) +O(\beta
^2)
\label{lagorderbeta}\eeq
with the momenta,
\beq \pi_\mu^x=0~,~\pi_\mu^p=-x_\mu(1-\frac{1}{2}\beta p^2)+\beta(x p)p_\mu
\label{pimu-orderbeta}\eeq
leading to the constraints,
\beq \phi_\mu^1=\pi_\mu^x\approx
0~,~\phi_\mu^2=\pi_\mu^p+x_\mu(1-\frac{1}{2}\beta p^2)-\beta(x p)p_\mu
\approx 0. \label{phi12orderbeta}\eeq
We find the Dirac Brackets to be, (with details in Appendix A),
\beq \{x^\mu,p^\nu\}=-\left[\frac{g_{\mu\nu}}{\left(1-\frac{\beta
p^2}{2}\right)}+\frac{\beta p_\mu p_\nu}{\left(1-\frac{3\beta
p^2}{2}\right)\left(1-\frac{\beta p^2}{2}\right)}\right], $$$$
 \{x^\mu,x^\nu\}=\{p^\mu,p^\nu\}=0. \label{db2} \eeq
Notice that the algebra is still structurally similar as the exact one and the
Snyder form with
non-zero $\{x_\mu,x_\nu\}$ has not appeared. This agrees with previous results
that the Snyder form
is present only in $O(\beta^2)$ or when more than one $\beta$-like parameters
are
present \cite{kem1,q}. However,
linearizing this algebra to $O(\beta )$ is once again problematic as it clashes
with the Jacobi identity. We will
see that the Snyder form is necessary in the linearized system in order to
exactly
satisfy the Jacobi identity.

The combination $x_\mu, (1-\frac{\beta p^2}{2})p_\nu $ constitute a canonical
pair, $\{x_\mu, (1-\frac{\beta p^2}{2})p_\nu\}=-g_{\mu\nu}$. The operator
$j_{\mu\nu}=(1-\frac{\beta p^2}{2})(x_\mu p_\nu -x_\nu p_\mu )$ transforms
$x_\mu $ and $p_\mu$ correctly and satisfies the correct Lorentz algebra
(\ref{lor}).
\vskip.4cm
{\bf{$O(\beta ^2 )$ {\it{results}}:}}
With  $\Lambda \approx 1+\frac{\beta p^2}{2}-\left(\frac{\beta p^2}{2}\right)^2$
the Lagrangian $L_{(2)}$ (without the mass-shell condition) becomes,
\beq L_{(2)}=-(x\dot{p})\left(1-\frac{\beta p^2}{2}+\left(\frac{\beta
p^2}{2}\right)^2\right)+\beta(x p)(p\dot{p})\left(1-\frac{3\beta p^2}{2}\right)
\label{l2} \eeq
yielding the constraints,
\beq \phi^1_\mu=\Pi^x_\mu,~\phi^2_\mu=\Pi^p_\mu+x_\mu\left(1-\frac{\beta
p^2}{2}+\left(\frac{\beta p^2}{2}\right)^2\right)-\beta p_\mu(x
p)\left(1-\frac{3\beta p^2}{2}\right). \label{c2} \eeq
The Dirac Brackets are, (with details in Appendix B),
\beq \{x_\mu,x_\nu\}=D(x_\mu p_\nu-x_\nu p_\mu),~ \{p_\mu,p_\nu\}=0 ,$$$$
\{x_\mu,p_\nu\}=-\frac{g_{\mu\nu}}{\left(1-\frac{\beta
p^2}{2}+\left(\frac{\beta p^2}{2}\right)^2\right)}-C p_\mu p_\nu \label{dbdb}
\eeq
 where $$C=\frac{\beta\left(1-\frac{3\beta p^2}{2}\right)}{\left(1-\frac{3\beta
p^2}{2}+\frac{7\beta^2 p^4}{4}\right)\left(1-\frac{\beta p^2}{2}+\frac{\beta^2
p^4}{4}\right)},~~D=\frac{C\beta p^2}{2\left(1-\frac{3\beta p^2}{2}\right)}.$$
We notice that the Snyder form has been recovered once $O(\beta ^2)$
contributions are introduced. This GUP based algebra - Snyder algebra connection
constitutes the other major result.
It is possible to construct the deformed Poincare generators but the expressions
are quite involved and not very illuminating.
\vskip.4cm
{\bf{{\it{Two parameter}} ($\beta,~\beta '$) {\it{ results}}:}} We now provide a
considerably simpler
Lagrangian with two parameters $\beta$ and $\beta '$ that can induce the Snyder
algebra. Note that {\it{ab initio}} it would have been hard to guess this result
as well as the explicit
expressions for the algebra but
in our constraint framework this is quite straightforward. From the constraint
analysis that generates the Dirac Brackets it is clear that we need a
non-vanishing
$\{\phi_2^\mu, \phi_2^\nu \}$ to reproduce a non-vanishing
$\{x^\mu, x^\nu \}$ bracket. Let us go back to (\ref{lagorderbeta}) and Appendix
A. It is
now clear that
the two non-canonical terms in $L_{(1)}$ must have different $\beta$-factors to
produce
the desired effect. Hence we consider the Lagrangian $L_{(\beta,\beta ')}$
(without the mass-shell condition),
\beq L_{(\beta,\beta ')}=-(x\dot{p})\left(1-\frac{\beta p^2}{2}\right)+\beta'(x
p)(p\dot{p}).
\label{ll} \eeq
The constraints of the model are,
\beq \phi^1_\mu=\Pi^x_\mu,~\phi^2_\mu=\Pi^p_\mu+x_\mu\left(1-\frac{\beta
p^2}{2}\right)-\beta'(x p)p_\mu ,\label{cc1} \eeq
giving rise to the Dirac Brackets,
 (with details in Appendix C),
\beq \{x_\mu,x_\nu\}=D\frac{(\beta-\beta')}{\beta '}(x_\mu p_\nu-x_\nu
p_\mu),~\{p_\mu,p_\nu\}=0 ,$$$$
 \{x_\mu,p_\nu\}=-\frac{g_{\mu\nu}}{\left(1-\frac{\beta p^2}{2}\right)}-D
p_\mu p_\nu ,\label{ddb} \eeq
 where $$D=\frac{\beta '}{\left(1-\frac{\beta p^2}{2}-\beta'
p^2\right)\left(1-\frac{\beta p^2}{2}\right)}.$$
Clearly for $\beta=\beta'\rightarrow \{x_\mu,x_\nu\}=0$ leaving a GUP like
algebra. We have not shown  the deformed Poincare generators which are quite
complicated.

Incidentally there is a simple linear model with Snyder algebra obtainable from
the previous one by
putting $\beta '=0$ and considering terms up to $O(\beta )$ only. The Lagrangian
$L_{S}$
\beq L=-(x\dot{p})\left(1-\frac{\beta p^2}{2}\right)+\frac{e}{2}(p^2-m^2),
\label{ls} \eeq
yields the Snyder algebra,
\beq \{x_\mu,x_\nu\}=\beta\frac{(x_\mu p_\nu-x_\nu p_\mu)}{(1-\frac{\beta
p^2}{2})^2},~
\{x_\mu,p_\nu\}=-\frac{g_{\mu\nu}}{(1-\frac{\beta p^2}{2})},~
\{p_\mu,p_\nu\}=0 .\label{sdbs} \eeq
Interestingly, its' O($\beta $) linearized version,
\beq \{x_\mu,x_\nu\}=\beta(x_\mu p_\nu-x_\nu p_\mu),~
\{x_\mu,p_\nu\}=-g_{\mu\nu}\left(1+\frac{\beta p^2}{2}\right),~
\{p_\mu,p_\nu\}=0 .\label{dbs} \eeq
also satisfies the Jacobi identity. One can check that
 $\left(\frac{x_\mu}{(1+\frac{\beta p^2}{2})},p_\mu\right)$ constitutes a
canonical pair with $j_{\mu\nu}=\frac{1}{(1+\frac{\beta p^2}{2})}(x_\mu
p_\nu-x_\nu p_\mu)$ being the deformed Lorentz generator.
\section{GUP particle in external electromagnetic field}
We introduce minimally coupled $U(1)$ gauge interaction to the free GUP particle
Lagrangian  (\ref{lag}),
\begin{equation}
L = -\frac{(x\dot{p})}{\Lambda}+\frac{\beta(x
p)(p\dot{p})}{\Lambda^2\sqrt{1+2\beta
p^2}}+\lambda(f(p^2)-m^2)+e(A\dot{x}).
\label{lagramgian}
\end{equation}
The symplectic structure is changed and we need to compute the new Dirac
algebra. 
The conjugate momentu follows are (\ref{lagramgian})
are
\begin{equation}
\pi^x_\mu=\frac{\partial L}{\partial
\dot{x}_\mu}=eA_\mu(x)~,~\pi^p_\mu=\frac{\partial L}{\partial
\dot{p}_\mu}=-\frac{x_\mu}{\Lambda}+
\frac{\beta(x p)p_\mu}{\Lambda^2\sqrt{1+2\beta p^2}}.\label{piXpiP}
\end{equation}
The constraints follow:
\begin{equation}
\phi^1_\mu=\pi^x_\mu-eA_\mu(x)~,~\phi^2_\mu=\pi^p_\mu+\frac{x_\mu}{\Lambda}
-\frac{\beta(x p)p_\mu}{\Lambda^2\sqrt{1+2\beta p^2}}.\label{phiXphiP}
\end{equation}
Since we have the Poission brackets
\begin{equation}
\{x_\mu,\pi^x_\nu\}=-g_{\mu\nu}~,~\{p_\mu,\pi^p_\nu\}=-g_{\mu\nu},
\end{equation}
the  constraint algebra is,
\begin{equation}
\{\phi^1_\mu,\phi^1_\nu\}=\{\pi^x_\mu-eA_\mu(x),\pi^x_\nu-eA_\nu(x)\}=-eF_{
\mu\nu}(x),\label{phi1phi1}
\end{equation}
\begin{eqnarray}
\{\phi^1_\mu,\phi^2_\nu\}=\{\pi^x_\mu-eA_\mu(x),\pi^p_\nu+\frac{x_\nu}{\Lambda}
-\frac{\beta(x p)p_\nu}{\Lambda^2\sqrt{1+2\beta p^2}}\}
=\frac{1}{\Lambda}g_{\mu\nu}-\frac{\beta p_\mu p_\nu}{\Lambda^2\sqrt{1+2\beta
p^2}},\label{phi1phi2}
\end{eqnarray}
\begin{equation}
\{\phi^2_\mu,\phi^2_\nu\}=\{\pi^p_\mu+\frac{x_\mu}{\Lambda}-\frac{\beta(x
p)p_\mu}{\Lambda^2\sqrt{1+2\beta p^2}},\pi^p_\nu
+\frac{x_\nu}{\Lambda}-\frac{\beta(x p)p_\nu}{\Lambda^2\sqrt{1+2\beta
p^2}}\}=0.\label{phi2phi2}
\end{equation}
This gives the constraint matrix as
\begin{eqnarray}
  \{\phi_\mu^i,\phi_\nu^j\} &=& \left[ {\begin{array}{cc}
 -e F_{\mu\nu} & \frac{g_{\mu\nu}}{\Lambda}-\frac{\beta p_\mu
p_\nu}{\Lambda^2\sqrt{1+2\beta p^2}}  \\
 -\frac{g_{\mu\nu}}{\Lambda}+\frac{\beta p_\mu p_\nu}{\Lambda^2\sqrt{1+2\beta
p^2}} & 0  \\
 \end{array} } \right] \nonumber\\
 &=& A+eB \label{phimuphinu}
 \end{eqnarray}
 where
\begin{equation}
A=\left[{\begin{array}{cc}
 0 & \frac{g_{\mu\nu}}{\Lambda}-\frac{\beta p_\mu p_\nu}{\Lambda^2\sqrt{1+2\beta
p^2}}  \\
 -\frac{g_{\mu\nu}}{\Lambda}+\frac{\beta p_\mu p_\nu}{\Lambda^2\sqrt{1+2\beta
p^2}} & 0  \\
 \end{array} } \right]
 ~~,~
B=\left[ {\begin{array}{cc}
 -F_{\mu\nu} & 0  \\
 0 & 0  \\
 \end{array} }\right] \label{matrixB}
\end{equation}
To $O(e)$ the inverse is $$(A+eB)^{-1}=A^{-1}-eA^{-1}BA^{-1}. $$
In the present case, the inverse of (\ref{phimuphinu}) to first order of $e$ is
\begin{equation}
\{\phi_\mu^i,\phi_\nu^j\}^{-1}=\left[ {\begin{array}{cc}
0 & -\Lambda g_{\mu\nu}-\beta p_\mu p_\nu  \\
\Lambda g_{\mu\nu}+\beta p_\mu p_\nu & -e\Lambda(\Lambda F_{\mu\nu}+\beta
p_a(F_{\mu a}p_\nu-F_{\nu a}p_\mu)) \\
\end{array} } \right]. \label{inv-phimuphinu}
\end{equation}
Therefore the Dirac brackets, modified by the $U(1)$ interaction,  are
\begin{eqnarray}
\{x_\alpha,x_\gamma\}^*=\{x_\alpha,x_\gamma\}-\{x_\alpha,\phi^1_\mu\}\{
\phi^1_\mu,\phi^1_\nu\}^{-1}\{\phi^1_\nu,x_\gamma\}=0, \label{diracXX}
\end{eqnarray}
\begin{eqnarray}
\{x_\alpha,p_\gamma\}^*=\{x_\alpha,p_\gamma\}-\{x_\alpha,\phi^1_\mu\}\{
\phi^1_\mu,\phi^2_\nu\}^{-1}\{\phi^2_\nu,p_\gamma\}
=-(\Lambda g_{\alpha\gamma}+\beta p_\alpha p_\gamma), \label{diracXP}
\end{eqnarray}
\begin{eqnarray}
\{p_\alpha,p_\gamma\}^*=\{p_\alpha,p_\gamma\}-\{p_\alpha,\phi^2_\mu\}\{
\phi^2_\mu,\phi^2_\nu\}^{-1}\{\phi^2_\nu,p_\gamma\}
=-e\Lambda(\Lambda F_{\alpha\gamma}+\beta p_a(F_{\alpha a}p_\gamma-F_{\gamma
a}p_\alpha)) .\label{diracPP}.
\end{eqnarray}
The relativistic Hamiltonian is,
\begin{equation}
H=\frac{p^2}{m}-\sqrt{p^2}\label{rel-ham}.
\end{equation}
Using the Dirac brackets (\ref{diracXX}-\ref{diracPP}) and (\ref{rel-ham}), the
Hamiltonian equations of motion are,
\begin{eqnarray}
\dot{x}_\alpha=\{x_\alpha,\frac{p_\gamma p_\gamma}{m}-\sqrt{p_\gamma
p_\gamma}\}^*
=-\frac{1}{m}\Lambda^2\sqrt{1+2\beta p^2}~p_\alpha, \label{dotx}
\end{eqnarray}
and
\begin{eqnarray}
\dot{p}_\alpha=\{p_\alpha,\frac{p_\gamma p_\gamma}{m}-\sqrt{p_\gamma
p_\gamma}\}^*
=-\frac{e}{m}\Lambda^2\sqrt{1+2\beta p^2}~p_\gamma F_{\alpha\gamma}.
\label{dotp}
\end{eqnarray}
Keeping only $O(e)$ terms we can eliminate $p$, to get the modified Newton's
law,
\begin{equation}
\ddot{x}_\alpha=-\frac{e}{m}\Lambda^2\sqrt{1+2\beta p^2}~\dot{x}_\gamma
F_{\alpha\gamma}=\frac{e}{m}\Lambda^2\sqrt{1+2\beta m^2}~\dot{x}_\gamma
F_{\alpha\gamma}.\label{equmotion}
\end{equation}
It is important to note that the dynamics in (\ref{dotp}) and (\ref{equmotion})
is {\it{exact}} forthe GUP parameter $\beta$ although it is  to the first order
of $e$. Hence the dynamics remains qualitatively unchanged with a
renormalization of the charge.
The  $O(\beta) $  equation of motion is
\begin{equation}
\ddot{x}_\alpha=-\frac{e}{m}(1+2\beta m^2)\dot{x}_\gamma F_{\alpha\gamma}.
\label{equmotion2}
\end{equation} 
\section{Summary and Conclusion}
The present paper deals with an extension of the Generalized Uncertainty
Principle in a covariant
setting. We have established a Lagrangian (or symplectic) framework where the
Generalized Uncertainty Principle
and extended Snyder Algebra are studied in a unified way. The above
non-commutative structures appear as
Dirac Brackets. It is instructive to see how the Lagrangian model for
Generalized Uncertainty Principle reduces to
the Snyder model at different level of approximation. Quite interestingly, the
more complicated Snyder algebra,
(since the coordinates do not commute among themselves), emerges from an
approximation of the model
for the simpler algebra compatible to the Generalized Uncertainty Principle. Our
analysis explicitly shows that, while considering  approximations,
it is not always pertinent to truncate a symplectic algebra directly as this
approximation
may invalidate Jacobi identities.
On the other hand it is always legitimate to start from a Lagrangian (or
symplectic structure), approximate at this level and subsequently compute the
algebra as Dirac Brackets, especially
when non-linear constraints are involved. But this requires construction of the
Lagrangian for a given Non-Commutative algebra as we have done here for the 
algebra based on Generalized Uncertainty
Principle. These generalized particle models and the connection between
Generalized Uncertainty
Principle based algebra and Snyder algebra are new results.

Finally we have studied behavior of the GUP particle in presence of external
$U(1)$ gauge interaction. We find that to lowest order
in $e$ (the gauge coupling) and to all orders in $\beta$ (Non-Commutative GUP
parameter), the charge gets modified without
any qualitative change in dynamics.

Apart from the above algebraic consistency arguments in favor of a Lagrangian
framework, from the physics
point of view it is indeed appealing and worthwhile to have a point particle
picturisation of the
noncommutative algebra in question as it clearly shows how it differs from the
conventional particle.

\vskip .5cm
\section{Appendices:}
In the Dirac Hamiltonian scheme of constraint analysis, non-commuting
constraints are termed as Second Class
Constraints. Consider a set of Second Class Constraints $\phi^i_\mu \approx 0$
where the constraint
commutator matrix $\{\phi^i_\mu,\phi^j_\nu \}=\Gamma^{ij}_{\mu\nu}$ in
non-singular with an inverse
$\Gamma_{ij}^{\mu\nu}$ such that
$$\Gamma^{ij}_{\mu\nu}\Gamma_{jk}^{\nu\lambda}=\delta^i_j\delta^\mu_\lambda .$$
The Dirac Bracket between two generic variables is  defined as,
\beq
\{A,B\}_{Dirac~ Bracket}= \{A,B\}-\{A,\phi^i\}\{\phi_i,\phi_j \}\{\phi^j,B\}
=\{A,B\}-\{A,\phi^i\}\Gamma_{ij}\{\phi^j,B\}.
\label{dira}
\eeq
This makes $\{\phi^i,A\}_{Dirac~Bracket}=0$ and
one can exploit $\phi^i=0$ strongly provided one uses Dirac Brackets. Upon
quantization the Dirac Brackets are
elevated to quantum commutators. Throughout our work we have dropped the
subscript $Dirac~Bracket$.\\
\textbf{Appendix A:} In the linearized model,  $\Lambda=\frac{\beta
p^2}{\sqrt{1+2\beta
p^2}-1}\simeq 1+\frac{1}{2}\beta p^2$ and
\beq L{(1)}\cong-(x\dot{p})\left(1-\frac{\beta p^2}{2}\right)+\beta(x
p)(p\dot{p}). \label{} \eeq
The momenta $ \Pi^x_\mu=0,~\Pi^p_\mu=-x_\mu\left(1-\frac{\beta
p^2}{2}\right)+\beta(x p)p_\mu $ gives the constraints $
\phi^1_\mu=\Pi^x_\mu,~\phi^2_\mu=\Pi^p_\mu+x_\mu\left(1-\frac{\beta
p^2}{2}\right)-\beta(x p)p_\mu $. The  constraint matrix
\begin{equation}
\Gamma^{ij}_{\mu\nu}=
\left[ {\begin{array}{cc}
 0 & g_{\mu\nu}\left(1-\frac{\beta p^2}{2}\right)-\beta p_\mu p_\nu \\
 -g_{\mu\nu}\left(1-\frac{\beta p^2}{2}\right)+\beta p_\mu p_\nu & 0 \\
 \end{array} }\right] \label{}
\end{equation}
has the inverse,
\begin{equation}
\Gamma_{ij\mu\nu}=
\left[ {\begin{array}{cc}
 0 & -\left(\frac{g_{\mu\nu}}{\left(1-\frac{\beta p^2}{2}\right)}+\frac{\beta
p_\mu p_\nu}{\left(1-\frac{3\beta p^2}{2}\right)\left(1-\frac{\beta
p^2}{2}\right)}\right) \\
 \left(\frac{g_{\mu\nu}}{\left(1-\frac{\beta p^2}{2}\right)}+\frac{\beta p_\mu
p_\nu}{\left(1-\frac{3\beta p^2}{2}\right)\left(1-\frac{\beta
p^2}{2}\right)}\right) & 0 \\
 \end{array} }\right].
\end{equation}
Using the definition  given above in (\ref{dira}) we find the Dirac Bracket
\beq \{x^\mu,p^\nu\}=-\left[\frac{g_{\mu\nu}}{\left(1-\frac{\beta
p^2}{2}\right)}+\frac{\beta p_\mu p_\nu}{\left(1-\frac{3\beta
p^2}{2}\right)\left(1-\frac{\beta p^2}{2}\right)}\right],~
\{x^\mu,x^\nu\}=\{p^\mu,p^\nu\}=0. \eeq

\textbf{Appendix B:} For $O(\beta^2)$ contribution,
$\Lambda=1+\frac{\beta p^2}{2}-\left(\frac{\beta p^2}{2}\right)^2$,

\beq L_{(2)}=-(x\dot{p})\left(1-\frac{\beta p^2}{2}+\left(\frac{\beta
p^2}{2}\right)^2\right)+\beta(x p)(p\dot{p})\left(1-\frac{3\beta p^2}{2}\right)
\eeq
with the constraints,
\beq \phi^1_\mu=\Pi^x_\mu,~\phi^2_\mu=\Pi^p_\mu+x_\mu\left(1-\frac{\beta
p^2}{2}+\left(\frac{\beta p^2}{2}\right)^2\right)-\beta p_\mu(x
p)\left(1-\frac{3\beta p^2}{2}\right). \eeq
The constraint matrix is
\begin{equation}
\Gamma^{ij}_{\mu\nu}=\{\phi^\mu_i,\phi^\nu_j\}$$$$
=\left[ {\begin{array}{cc}
 0 & g_{\mu\nu}\left(1-\frac{\beta p^2}{2}+\left(\frac{\beta
p^2}{2}\right)^2\right)-\beta p_\mu p_\nu\left(1-\frac{3\beta p^2}{2}\right) \\
 -g_{\mu\nu}\left(1-\frac{\beta p^2}{2}+\left(\frac{\beta
p^2}{2}\right)^2\right)+\beta p_\mu p_\nu\left(1-\frac{3\beta p^2}{2}\right) &
\frac{1}{2}\beta^2 p^2(x_\mu p_\nu-x_\nu p_\mu) \\
 \end{array} }\right]
\end{equation}
has the inverse,
\begin{equation}
\Gamma_{ij\mu\nu}=\{\phi^\mu_i,\phi^\nu_j\}^{-1}=
\left[ {\begin{array}{cc}
  D(x_\mu p_\nu-x_\nu p_\mu) & -\frac{g_{\mu\nu}}{\left(1-\frac{\beta
p^2}{2}+\left(\frac{\beta p^2}{2}\right)^2\right)}-C p_\mu p_\nu \\
 \frac{g_{\mu\nu}}{\left(1-\frac{\beta p^2}{2}+\left(\frac{\beta
p^2}{2}\right)^2\right)}+C p_\mu p_\nu & 0 \\
 \end{array} }\right]
\end{equation}
where $$C=\frac{\beta\left(1-\frac{3\beta p^2}{2}\right)}{\left(1-\frac{3\beta
p^2}{2}+\frac{7\beta^2 p^4}{4}\right)\left(1-\frac{\beta p^2}{2}+\frac{\beta^2
p^4}{4}\right)}$$\\
and $$D=\frac{C\beta p^2}{2\left(1-\frac{3\beta p^2}{2}\right)}.$$
Then the Dirac Brackets are
\beq \{x_\mu,x_\nu\}=D(x_\mu p_\nu-x_\nu p_\mu),~\{p_\mu,p_\nu\}=0 ,~
 \{x_\mu,p_\nu\}=-\frac{g_{\mu\nu}}{\left(1-\frac{\beta
p^2}{2}+\left(\frac{\beta p^2}{2}\right)^2\right)}-C p_\mu p_\nu. \eeq
\textbf{Appendix C:} For two parameters $\beta$ and $\beta'$ we have
\beq L_{(\beta,\beta ')}=-(x\dot{p})\left(1-\frac{\beta p^2}{2}\right)+\beta'(x
p)(p\dot{p}) \eeq
The constraints are
\beq \phi^1_\mu=\Pi^x_\mu,~\phi^2_\mu=\Pi^p_\mu+x_\mu\left(1-\frac{\beta
p^2}{2}\right)-\beta'(x p)p_\mu \eeq
The constraint matrix is
\begin{equation}
\Gamma^{ij}_{\mu\nu}=\{\phi^\mu_i,\phi^\nu_j\}=
\left[ {\begin{array}{cc}
  0 & g_{\mu\nu}\left(1-\frac{\beta p^2}{2}\right)-\beta'p_\mu p_\nu \\
 -g_{\mu\nu}\left(1-\frac{\beta p^2}{2}\right)+\beta'p_\mu p_\nu &
(\beta-\beta')(x_\mu p_\nu-x_\nu p_\mu) \\
 \end{array} }\right]
\end{equation}
with the inverse,
\begin{equation}
\Gamma_{ij\mu\nu}=\{\phi^\mu_i,\phi^\nu_j\}^{-1}=
\left[ {\begin{array}{cc}
  \frac{D}{\beta'}(\beta-\beta')(x_\mu p_\nu-x_\nu p_\mu) &
  -\frac{g_{\mu\nu}}{\left(1-\frac{\beta p^2}{2}\right)}-D p_\mu p_\nu \\
 \frac{g_{\mu\nu}}{\left(1-\frac{\beta p^2}{2}\right)}+D p_\mu p_\nu & 0 \\
 \end{array} }\right],
\end{equation}
where $$D=\frac{\beta'}{\left(1-\frac{\beta p^2}{2}-\beta'
p^2\right)\left(1-\frac{\beta p^2}{2}\right)}.$$
The Dirac Brackets are
\beq \{x_\mu,x_\nu\}=\frac{D}{\beta'}(\beta-\beta')(x_\mu p_\nu-x_\nu p_\mu),
~\{p_\mu,p_\nu\}=0 ,
~\{x_\mu,p_\nu\}=-\frac{g_{\mu\nu}}{\left(1-\frac{\beta p^2}{2}\right)}-D p_\mu
p_\nu. \eeq

\end{document}